\documentclass[11pt,a4paper]{scrartcl}

\usepackage{CLICdp}

\usepackage{CLICdp_definitions}

\title{SiD Status and Plans}

\date{\formatdate{31}{3}{2017}}

\addauthor{A.~Robson}{\institute{3}}

\addinstitute{3}{University of Glasgow, UK}

\onbehalfof{the SiD Consortium}

\abstract{Recent work carried out in the SiD Consortium is reported. Results have been obtained with the Chronopix version 3 chip, intended for the SiD vertex detector.  A test structure is being produced for KPiXM, a CMOS MAPS approach that could be used for the tracker and electromagnetic calorimeter.  Tracker sensors are being tested and a new prototype sensor is being planned.  A new effort on tracker support structures has begun and first prototypes produced.  New studies of the pair background envelope in the beam pipe have been carried out.  Electromagnetic calorimeter testbeam data has been analysed and plans are being developed for a full stack prototype.  New comprehensive studies of high-rate backgrounds and muons from the beam delivery system have been carried out to determine aspects of the forward region layout. Work has started on alignment and calibration strategies; and progress has been made on a new simulation and reconstruction framework for SiD.
}

\titlecomment{Talk presented at the International Workshop on Future Linear Colliders (LCWS2016)\\
		Morioka, Japan, 5-9 December 2016\\
		C16-12-05.4 \\ \vspace*{5mm}
Slides available at: https://agenda.linearcollider.org/event/7371/contributions/37654/ }

\notitlestamp

\graphicspath{ {./logos/}{./figures/} }

\begin{document}

\titlepage

\section{Introduction}

SiD [1,2] is a compact, cost-constrained, general-purpose detector proposed for 
the International Linear Collider: a future e$^+$e$^{-}$ collider 
with baseline energy of $\sqrt{s}=500$\,GeV and potential to operate 
up to $\sqrt{s}=1$\,TeV.
This contribution reports the SiD Consortium's R\&D and software developments
since the 2015 International Workshop on Future Linear Colliders.

\section{SiD Overview}

SiD is designed to make precision measurements and to be sensitive to a wide 
range of new phenomena.
The detector consists of several subsystems.
The sensitive layers of the vertexing and tracking system are entirely silicon 
and provide excellent momentum resolution and single bunch-crossing 
time-stamping.
Highly segmented calorimeters optimized for particle flow sit inside a 
solenoid of inner radius 2.6\,m that provides a 5\,T $B$-field.  
The coil is surrounded by an iron flux return that incorporates muon chambers,
and the complete assembly is mounted on a platform designed for rapid 
push-pull operation.
Further details on the subsystems and on recent work 
are given in the following sections.

\section{Vertexing and Tracking}

The vertex detector consists of five short barrel layers starting 
at radius $r=1.4$\,cm, and five disks.  The required hit 
resolution is better than 3\,$\mu$m and the system needs to provide 
single-bunch timing resolution, and keep within a material budget of 
around 0.1\% $X_0$ per layer, and a power budget of less than 
130\,$\mu$W/mm$^2$.
As the vertex detector will be one of the final systems to be installed, 
the technology choice for implementation can be made later; options 
include `standard' silicon diode pixels, monolithic designs such as 
Chronopix, vertically integrated chips, or high-voltage CMOS.

The current baseline readout for the vertex detector is the Chronopix chip.
Its concept has been proven through a series of three prototypes [3]. 
The first version demonstrated 300\,ns timestamping, sparse readout, 
and pulsed powering; the second version demonstrated NMOS electronics 
with acceptable power consumption and comparator offset calibration; 
and the third, current, version has six sensor options to address 
a large sensor capacitance that was observed in version 2 and 
appears now to be solved.  Cross-talk issues have also been addressed 
by separating analogue and digital power and adding a decoupling 
capacitor.  Ongoing work is to fully characterise the sensor 
operation: sensor efficiency for MIPs and radiation hardness 
measurements are in progress with the prototype 3 chips.

The silicon strip tracker extends to $r=1.25$\,m and consists 
of five barrel layers and four disks of 25\,$\mu$m pitch strips 
with 50\,$\mu$m readout.  The system is gas-cooled and has a 
total material budget of less than 20\% $X_0$ in the active region.

The current baseline readout for the tracker uses the KPiX ASIC [4], 
bump-bonded to 
the module.  There is also ongoing development work on KPiXM, a 
CMOS MAPS approach where the sensors and front-end electronics are 
integrated on the same substrate.  This has potential for a lower 
material budget, smaller pixel size that is not limited by bump-bonding, 
and lower cost through implementation in standard commercial technologies.
A test structure is currently being produced in 150\,nm technology, 
using a high-resistivity substrate thinned to 150\,$\mu$m.  Several 
variants of active and passive pixels of $40\times 500\,\mu$m$^2$
are included.  A variant with larger $1000\times 1000\,\mu$m$^2$ 
pixels could be used for the SiD electromagnetic calorimeter.

There has been progress on the tracker sensors.  A previous Hamamatsu 
prototype was found to be damaged when undergoing wirebonding. 
Recently, agreement has been 
reached to produce a new prototype with increased oxide layer thickness 
between the two metal layers to address 
this problem, as well as under-bump metallization. 
 The Consortium is moving towards a full 
prototype test consisting of sensor, KPiX readout, and attached cables.

Furthermore, there has been new effort on tracker support structures 
by groups in the UK.  This aims to build structures that integrate 
services and cooling, with lengths of several metres and material 
budget less than 1\% $X_0$, using carbon fibre 
reinforced polymer box channels.  Services can be co-cured into the 
structure, and adjacent box channels linked by tongue-and-groove joints.  
Tracker modules would sit on both sides of the resulting hoop structures.  
FEA studies have been done, and first prototypes produced.

Finally, there have been updated studies of 
the pair background envelope in the beampipe.  These are described 
in detail elsewhere in these proceedings [5]; in summary: with the 
current beampipe design, only around 0.45\% of all particles leave 
tracks outside the beampipe.  A reduction in the beampipe radius of 
up to 2\,mm could therefore be considered, and even the addition of an 
additional vertex detector layer; however, this would require further 
studies of sychrotron radiation and flavour tagging performance.

\section{Calorimetry}

A highly granular `imaging' calorimeter is essential for the ILC 
physics programme as an integral part of the particle flow reconstruction 
approach.

The baseline design for the electromagnetic calorimeter consists of 
20 thin and 10 thick tungsten layers interspersed with silicon sensitive  
layers.  The silicon sensors are hexagonal arrays of 1024 pads, each 
13\,mm$^2$, bump-bonded to a KPiX ASIC.  Experience so far with 
test sensors has shown that bump bonding to sensors with aluminium 
pads can be very difficult; under-bump metallization should help with 
that.  In addition, traces in the metal-2 layer 
from pixels to the pad array that run over other pixels 
have been observed to cause parasitic capacitance and cross-talk; 
in an updated design, a fixed potential trace in the metal-1 layer 
shields the signal traces from the pixels.
New prototypes with KPiX attached are currently under preparation 
for cable attachment.

A nine-layer prototype corresponding to around 6\,$X_0$ has been 
tested in a 12\,GeV electron testbeam at SLAC.  Details of the analysis 
are given elsewhere in these proceedings [6], showing good identification 
of showers that are separated by at least 1\,cm in the calorimeter.
Plans are being developed for a full stack prototype.

The Detector Baseline Design for the hadronic calorimeter consisted 
of digital readout using resistive plate chambers 
with $1\times 1$\,cm$^2$ tiles.  Following collaboration-wide review, 
the baseline techology has been updated to analogue readout 
using scintillator 
and silicon photomultipliers, with $3\times 3$\,cm$^2$ tiles.  
The hadronic calorimeter is composed of 40 layers.  
Work is ongoing to compare single-particle energy resolutions 
with CALICE testbeam results, and further work on the mechanical 
design is foreseen following the rebaselining.

\section{Solenoid and Muon System}
A recent redesign has introduced a 30$^\circ$ junction between 
the barrel and doors of the return yoke.
This has the benefit of more efficient flux return and lower 
stray fields, and also allowing easier transport and handling by 
redistributing the material between the barrel and doors.

The SiD baseline has the muon system consisting of long 
scintillator strips with wavelength-shifting fibre and silicon 
photomultiplier readout; this is a consistent extension of the 
updated baseline hadronic calorimeter technology.  Further 
optimization of the strip dimensions and number of layers will 
be carried out.  Since the Detector Baseline Document, the 
yoke and muon system has been updated from an eightfold to 
twelve-fold geometry to match the calorimeters.

\section{Forward Region Layout}
A recent comprehensive set of studies considering high cross-section 
processes such as pair-production arising from beam-beam interactions, 
Bhabha scattering, and two-photon 
processes, has addressed several aspects of the forward region layout.  
These are described in detail in a dedicated report [7].  

Systematic studies were done of the shape of the BeamCal and the effect of 
an anti-DID (Detector-Integrated Dipole) on the inner detector occupancy.
Studies were carried out using the $\sqrt{s}=500$\,GeV luminosity upgrade 
beam parameters in order to be conservative. 
The vertex detector barrel occupancy was shown to be 
robust under different BeamCal geometry variants and with or without 
the anti-DID. Furthermore it was seen that background pairs can arrive 
as long as microseconds after the beam crossing, which allows the 
possibility of rejecting them using timing cuts.

The forward calorimeter occupancies were also studied in order to 
determine the necessary buffer depth.  With a depth of 4 buffers, 
around 10$^{-3}$ of hits were lost; with a depth of 6 buffers this 
was reduced to 10$^{-4}$ of hits; however, in order to maintain 
losses below 10$^{-4}$ of hits for the innermost radii, 8 buffers 
were found to be needed.

Finally, the effect of muons from the beam delivery system was studied, 
as described elsewhere in these proceedings [8].
Magnetized spoilers are intended to sweep muons from the beam delivery 
system into the tunnel walls.  Full simulation of the detector found 
the silicon tracker endcap occupancy to be acceptable with these 
spoilers.  However, the electromagnetic calorimeter endcap occupancy 
was very high; this could be reduced by introducing a magnetized wall in 
addition to the spoilers.

\section{Simulation and Reconstruction}
At the 2015 International Workshop on Linear Colliders, SiD 
decided to implement its simulation in DD4hep [9], and adopt 
a common reconstruction in order to benefit from shared 
effort among the different linear collider detector concepts. 
Since then there has been a lot of progress in implementing 
the geometry, drivers, and digitizers and developing performance 
benchmarking tools.  Currently, tracking pattern recognition 
and Pandora particle flow reconstruction are being commissioned.

\section{Alignment and Calibration Strategies}
Track-based alignment will be essential for high-precision tracking 
at SiD.  Owing to the low cross-sections of relevant processes, 
there will be limited high-$p_T$ tracks to carry out the alignment. 
Studies indicate that around 1000 tracks per month per module in the 
outer tracker will be available during ramp-up in the first year, 
and SiD has no reason to believe that this would be significantly 
enhanced by running at $\sqrt{s}=91$\,GeV.
Different strategies, such as frequency-scanning interferometry, 
or modification of the electronics to increase efficiency for 
cosmic-ray tracks, will be investigated to contribute to the alignment.

\section{Outlook}
SiD is a compact, capable detector with a well-defined baseline that 
exceeds the physics requirements.  Nonetheless there is ongoing activity
in readout, sensors, structures, layout, and simulation and reconstruction
as described here.  
Over the coming year, among other things
 the SiD Consortium will continue to characterize 
the Chronopix sensor and new prototypes for KPiXM and a tracker sensor 
prototype; 
plans will be developed for a full stack electromagnetic calorimeter 
prototype; and work will continue on the new simulation and reconstruction 
chain. 
New members are welcome to join the SiD Consortium 
to contribute to this effort in preparation for 
the technical design phase. 

\section{Acknowledgements}
Thanks to the many members of the SiD Optimization Group who provided
input, including Marty Breidenbach, Anne Sch\"utz, Marcel Stanitzki, Jim Brau, Amanda Steinhebel, Bruce Schumm, Andy White, Ross McCoy, Andrew Myers, Georg Viehauser, Joel Goldstein, Dan Protopopescu, Bogdan Mischenko, and Jan Strube.

\section{References}
1. Aihara, H. et al., {\em SiD Letter of Intent}, 2009. arXiv:0911.0006\\
2. Behnke, T. et al., {\em The International Linear Collider Technical Design Report - Volume 4: Detectors}, 2013.  arXiv:1306.6329\\
3. Sinev, N., Brau, J., Strom, D., Baltay, C., Emmet, W., and Rabinowitz, D., {\em Chronopixel Project Status}, PoS(Vertex2015)038 (2015)\\
4. Brau, J., Breidenbach, M. et al., {\em KPiX - A 1024 Channel Readout ASIC for the ILC}, SLAC-PUB-15285 (2013), 2012 IEEE Nuclear Science Symposium\\
5. Sch\"utz, A., {\em Pair Background Envelopes in the SiD Detector}, arXiv:1703.05737\\
6. Steinhebel, A., and Brau, J., {\em Studies of the Response of the SiD Silicon-Tungsten ECal}, arXiv:1703.08605\\
7. Barklow, T., d'Hautuille, L., Milke, C., Schumm, B., Sch\"utz, A., Stanitzki, M., and Strube, J., {\em A Study of the Impact of High Cross Section ILC Processes on the SiD Detector Design}, 2016. arXiv:1609.07816\\
8. Sch\"utz, A., Keller, L., and White, G., {\em A Study of the Impact of Muons from the Beam Delivery System on the SiD Performance}, arXiv:1703.05738\\
9. Frank, M., Gaede, F., Grefe, C., and Mato, P., {\em DD4hep: A Detector Description Toolkit for High Energy Physics Experiments}, J. Phys. Conf. Series \textbf{513} (2014) 022010

\end{document}